\documentclass[sigconf]{acmart}
\usepackage{tcolorbox} 
\usepackage{multirow}
\usepackage{tabularx}
\usepackage{booktabs}
\usepackage{colortbl}
\usepackage{multirow}
\definecolor{lightgray}{gray}{0.90}
\newcommand{\blue}[1]{\textcolor{black}{#1}}

\AtBeginDocument{%
  }

\setcopyright{acmlicensed}

\copyrightyear{2026}
\acmYear{2026}
\setcopyright{cc}
\setcctype{by}
\acmConference[COMPASS '26]{ACM SIGCAS/SIGCHI Conference on Computing and Sustainable Societies}{July 27--31, 2026}{Virtual Event, USA}
\acmBooktitle{ACM SIGCAS/SIGCHI Conference on Computing and Sustainable Societies (COMPASS '26), July 27--31, 2026, Virtual Event, USA}
\acmDOI{10.1145/3811242.3819112}
\acmISBN{979-8-4007-2702-3/2026/07}

\begin{document}

\title{SpheriCity: Designing Trustworthy Conversational AI for Sustainability Decision Support}


\author{Ahmed Qayyum}
\affiliation{%
  \institution{Department of Computer Science, Colby College}
  \city{Waterville}
  \state{Maine}
  \country{USA}}
\email{ahmed.qayyum@colby.edu}

\author{Madison Werner}
\affiliation{%
  \institution{Circularity Informatics Lab, University of Georgia}
  \city{Athens}
  \state{Georgia}
  \country{USA}}
\email{madison.werner@uga.edu}

\author{Kathryn Youngblood}
\affiliation{%
  \institution{Circularity Informatics Lab, University of Georgia}
  \city{Athens}
  \state{Georgia}
  \country{USA}}
\email{kathryny@uga.edu}

\author{Jenna R. Jambeck}
\affiliation{%
  \institution{Circularity Informatics Lab, University of Georgia}
  \city{Athens}
  \state{Georgia}
  \country{USA}}
\email{jjambeck@uga.edu}

\author{Tahiya Chowdhury}
\affiliation{%
  \institution{Department of Computer Science, Colby College}
  \city{Waterville}
  \state{Maine}
  \country{USA}}
\email{tahiya.chowdhury@colby.edu}

\renewcommand{\shortauthors}{Qayyum et al.}

\begin{abstract}
We present \textbf{SpheriCity}, an expert-grounded conversational prototype designed to support trustworthy knowledge sensemaking from sustainability reports. City-level circularity assessment reports contain rich information about materials, infrastructure, and policy interventions, yet their length and heterogeneous structure make cross-document synthesis and comparison difficult for practitioners and researchers working on circular economy initiatives. While large language models (LLM) promise faster knowledge access and synthesis, their opaque reasoning, hallucinations, and lack of source transparency introduce risks for trust and interpretability, and require verification in high-stakes sustainability contexts.
SpheriCity addresses these challenges through a \textit{provenance-first}
conversational agent that foregrounds evidence traceability, structured synthesis, and interaction scaffolds to support exploratory querying and cross-document synthesis across sustainability reports. We conducted a formative expert review with six sustainability experts using representative queries spanning cross-city comparison, policy summarization, and recommendation-oriented tasks. Experts evaluated responses across dimensions, including relevance, contextual fit, accuracy, neutrality, and depth, and provided qualitative reflections on the system’s usefulness for sustainability knowledge work.
Our results reveal that transparent sourcing, contextual explanation, interpretability, and alignment with expert workflow strongly shape expert trust and judgments of system usefulness. This work contributes (1) a conversational prototype for sustainability knowledge sensemaking, (2) an expert-grounded evaluation framework for assessing AI responses in high-stakes knowledge domains, and (3) design insights into how provenance, uncertainty communication, and integration in workflow influence expert users' trust in AI assistance for sustainability decision support.
\end{abstract}

\ccsdesc[300]{Information systems~Expert systems}
\ccsdesc[500]{Human-centered computing~Interaction design}
\keywords{Sustainability, Information Retrieval, Decision Support Systems, Conversational AI.}

\received{20 February 2007}
\received[revised]{12 March 2009}
\received[accepted]{5 June 2009}

\maketitle


\section{Introduction}

Sustainability experts increasingly rely on large, heterogeneous report corpora to understand material flows, evaluate policy interventions, and compare practices across cities. City-level circularity assessment reports, environmental policy documents, and technical studies contain rich insights about infrastructure, materials management, and governance strategies to reduce waste leakage (e.g., single-use plastic) into
the environment and maximize the use of
resources~\cite{youngblood2022rapid, maddalene2023circularity}. However, this knowledge is often fragmented across hundreds of pages of reports, making cross-document synthesis difficult in real-world decision-making contexts that often operate under limited resource constraints. 

Consider a circular economy expert comparing plastic leakage mitigation strategies across coastal cities. To answer a single question, the expert may need to manually search and read multiple 80- to 100-page reports, identify relevant sections, cross-reference policies, and interpret dispersed evidence to generate actionable insights. This process imposes a substantial interactional and cognitive burden: important connections between knowledge fragments may be missed, data interpretation and comparison may be delayed, and policy decisions may be slowed by the time required to synthesize dispersed information from different sources. These reports are often used to inform policy design, urban planning strategies, and environmental interventions, making the ability to interpret and verify synthesized information a critical step for responsible sustainability decision-making.

Recent advances in large language models (LLMs) promise new possibilities for interacting with unstructured knowledge sources through conversational interfaces. Retrieval-augmented language models can synthesize information across documents and potentially accelerate knowledge access. However, in high-stakes sustainability contexts, these systems also introduce new challenges. Opaque synthesis processes, hallucinated facts, and missing source attribution can undermine user trust and make it difficult to assess the credibility of generated responses. As a result, conversational AI introduces a tension between speed and reliability: while these systems may accelerate information access, they may also amplify interpretability and trust risks that are particularly consequential in sustainability decision-making.

From an HCI perspective, this raises a central interaction challenge: \textbf{how can conversational agents be designed to support trustworthy knowledge sensemaking in sustainability contexts?} Prior work in sustainable HCI has emphasized the importance of designing computational systems that align with human values, accountability, and long-term societal impact~\cite{disalvo2010sustainable, knowles2013sustainability}. Research on human–AI collaboration and expert sensemaking similarly highlights the need for systems that make uncertainty visible, support verification, and calibrate user trust rather than replacing human judgment~\cite{amershi2019guidelines, lim2009intelligibility, muller2019datascience}. Yet, most LLM-based sustainability tools remain framed primarily as technical information retrieval systems, with limited attention to domain-relevant interaction design, interpretability, or how experts actually reason with AI-generated knowledge.

Prior work on LLM-powered chatbots has primarily focused on response generation and retrieval performance. Evaluations for such model-generated dialogue typically rely on human annotators rating responses along general-purpose dimensions such as fluency, grammatical correctness, and humanness~\cite{tian2019learning, li2018syntactically, moghe2018towards, ahn2022retrieval}, relevance~\cite{sai-etal-2020-improving, mehri-eskenazi-2020-usr}, and informativeness~\cite{liu2016evaluate, deriu2021survey}. In high-stakes domains such as sustainability, however, these generic evaluation criteria are insufficient. Experts must determine whether information is verifiable, appropriately scoped, unbiased, and traceable to primary evidence sources. Prior HCI research on conversational systems has explored design parameters for task-oriented interaction~\cite{clark2019conversation}, but has largely focused on everyday user tasks rather than expert knowledge synthesis and decision-relevant reasoning. To our knowledge, little prior work has examined how conversational AI systems can support sustainability experts working with large document collections where provenance, interpretability, and cross-source reasoning are central. This creates a gap between how conversational AI systems are generally evaluated and how experts actually judge trustworthiness in practice with their domain knowledge. 

To address this gap, we present \textbf{SpheriCity}, an expert-grounded conversational prototype designed to support trustworthy knowledge sensemaking over fragmented sustainability reports. SpheriCity takes a \textit{provenance-first} design that foregrounds evidence traceability through page-level source citations, structured response synthesis, and visible interaction cues that support exploratory querying. The system was developed through a participatory process with sustainability experts who co-developed a formative evaluation framework emphasizing five key dimensions: topical relevance, contextual fit, perceived accuracy, neutrality, and informational depth, reflecting the needs in sustainability decision making.

We then conducted a human expert evaluation with six sustainability practitioners using 13 representative queries spanning cross-city comparison, policy summarization, and recommendation-oriented tasks, resulting in a mixed-method approach to understand expert perceptions of trustworthiness, interpretability, and usefulness of conversational AI in sustainability knowledge workflows. Our findings highlight the importance of transparent sourcing, contextual explanation, and workflow alignment in shaping expert trust in conversational AI systems that can help design future decision support systems with AI assistance. \\

\textbf{Contributions.} In this work, we make the following contributions:

\begin{itemize}
    \item We present SpheriCity, a provenance-first conversational agent designed to support sustainability knowledge sensemaking across heterogeneous circularity assessment reports.

    \item We introduce an expert-grounded evaluation framework for assessing AI-provided sustainability knowledge, developed through participatory design with domain experts.

    \item Through a formative expert study, we identify interaction design principles for trustworthy conversational systems in high-stakes sustainability contexts, highlighting the importance of provenance, interpretability, and workflow alignment.
\end{itemize}

\blue{Importantly, our evaluation framework was applied through a static expert review protocol — experts rated pre-generated system outputs rather than interacting with the system live — enabling controlled, within-rater comparison across three retrieval strategies while reducing evaluation burden on expert users.} As our project \textit{SpheriCity} continues to evolve, this work presents the current prototype and formative findings to inform future studies of human–AI collaboration in sustainability knowledge work.


\begin{figure*}[t!]
\centering
\includegraphics[width=0.99\textwidth]{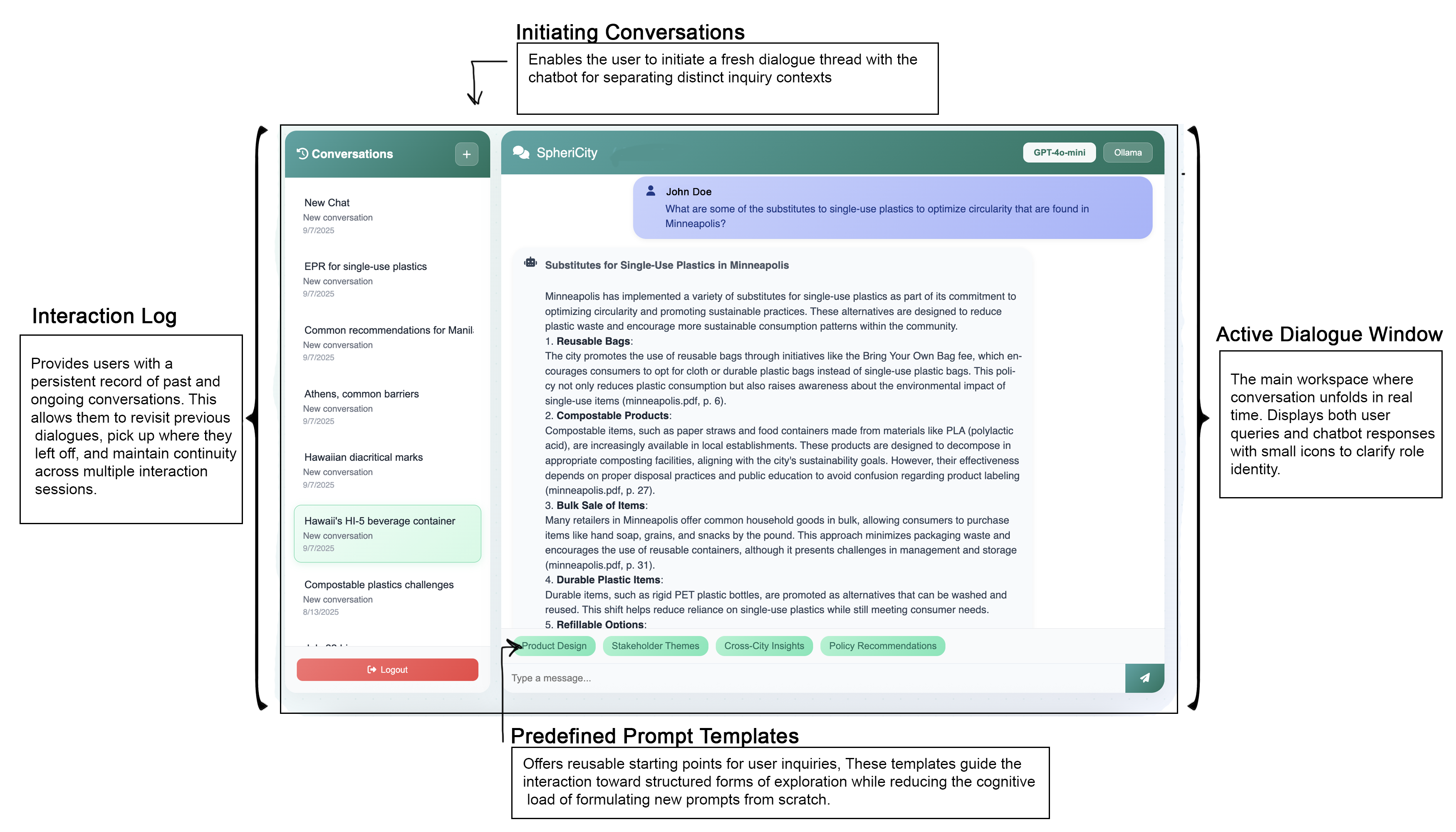}
\caption{Prototype chatbot interface for SpheriCity developed in this work. The interface includes (1) a conversational query panel for natural language questions and prompt templates, (2) a structured response display organizing generated insights, and (3) an evidence provenance panel linking responses to page-level citations from the sustainability reports.}
\Description{Prototype chatbot interface for SpheriCity developed in this work. The interface includes (1) a conversational query panel for natural language questions and prompt templates, (2) a structured response display organizing generated insights, and (3) an evidence provenance panel linking responses to page-level citations from the sustainability reports.}
\label{fig:chatbot}
\end{figure*}

\section{Related Work}

In this section, we review prior work related to three areas that are central to our study: (1) computing systems for sustainability knowledge and decision support,
(2) human-AI collaboration, and (3) evaluation of conversational AI systems.

\subsection{Computing for Sustainability and Knowledge Support}

Computing research has increasingly explored how digital systems can support sustainability transitions by enabling a better understanding of environmental systems, policy trade-offs, and resource management practices. Sustainable HCI research emphasizes the importance of designing computational tools that align with long-term societal values, support accountability, and promote responsible decision-making~\cite{disalvo2010sustainable, knowles2013sustainability, 10.1145/2493431}. Within sustainability practice, experts frequently rely on large document collections such as environmental reports, circular economy assessments, and policy analyses to understand material flows, evaluate interventions, and compare practices across regions. However, these knowledge sources are often fragmented across heterogeneous documents that are difficult to synthesize efficiently.

Recent work has explored how artificial intelligence and data-driven tools may support sustainability decision-making and the United Nations Sustainable Development Goals~\cite{vinuesa2020role}. These works explore conversational agents for environmental sustainability, including persuasive systems for pro-environmental behavior change, sustainable fashion guidance, and ecologically oriented community engagement. A number of recent review papers show that much of this literature focuses on persuasive conversational agents designed to shift everyday user behavior~\cite{giudici2025persuasive}, while other studies examine sustainability-oriented conversational support in consumer-facing domains such as fashion behavior~\cite{manzo2025sustainablefashion} or understanding ecological consideration beyond human perspectives~\cite{wong2024beyondhuman}. While this work establishes the relevance of conversational systems in sustainability-oriented HCI, it remains largely focused on behavior change, consumer guidance, or public engagement rather than expert-facing knowledge synthesis and decision support. Many existing tools focus primarily on predictive modeling or data analysis, rather than supporting experts in interpreting and synthesizing complex textual knowledge sources. As a result, we have a limited understanding of how conversational AI systems can support sustainability experts in navigating and interpreting report-based knowledge. 

Our work, on the other hand, examines how a conversational AI system can support sustainability experts working with large, heterogeneous report corpora where provenance, interpretability, and verification matter as much as answer usefulness. More broadly, this perspective aligns with calls for human-centric AI approaches to sustainable development and decision support~\cite{how2020humancentric}. 


\subsection{Human--AI Collaboration and Expert Sensemaking}

Research in human--AI interaction has increasingly emphasized the importance of designing systems that support human reasoning rather than replacing it. Studies of data science and knowledge work highlight that experts engage in iterative workflows involving exploration, interpretation, and verification of information sources~\cite{muller2019datascience}. In such contexts, AI systems must support human sensemaking by making underlying reasoning processes visible and enabling users to interrogate and validate system outputs.

Prior work has identified key design principles for trustworthy human--AI collaboration. Systems should provide intelligible explanations and surface evidence to support user interpretation~\cite{lim2009intelligibility}, expose uncertainty and limitations, and calibrate user trust appropriately~\cite{amershi2019guidelines}. More recent work further frames AI systems as collaborative partners rather than passive tools, highlighting the emergence of ``generative artificial experts'' that support decision-making through conversational interaction~\cite{sowa2025expert}. These systems extend earlier expert systems by enabling interactive, language-based access to complex knowledge sources.
In parallel, research on conversational interfaces for decision support has demonstrated their potential to enhance user engagement and trust in complex domains such as healthcare and environmental management~\cite{10.1145/3485447.3512248, shrestha2025community}. For example, conversational decision-support systems have been shown to improve users' ability to interpret data and make informed decisions by integrating evidence from multiple sources and presenting it through interactive dialogue. However, prior work also highlights that trust in such systems is highly sensitive to perceived accuracy, transparency, and the ability to verify underlying reasoning.

Within sustainability-focused HCI, conversational agents have primarily been explored as tools for influencing individual behavior change, such as promoting energy conservation or sustainable consumption practices~\cite{giudici2025persuasive}. These systems often emphasize persuasion, motivation, or engagement, rather than supporting expert analysis or decision-making workflows. Emerging work has begun to explore conversational AI for environmental data access and decision support~\cite{vald2024integrating, shrestha2025community}, but remains limited in addressing how experts critically evaluate, verify, and reason with AI-generated knowledge.

Despite these advances, most conversational AI systems still prioritize fluent response generation over interpretability, verification, and traceability of information sources. In high-stakes sustainability contexts, where decisions must be justified, contextualized, and accountable, these gaps become particularly consequential. Our work builds on this literature by explicitly designing for expert sensemaking, foregrounding provenance, and aligning conversational interaction with the verification and reasoning practices of sustainability professionals.

\subsection{Evaluation of Conversational AI Systems}

Evaluation of conversational AI systems has traditionally focused on response quality metrics such as fluency, grammatical correctness, and topical relevance. Prior work has shown that commonly used automatic metrics often correlate poorly with human judgments of response quality~\cite{liu2016evaluate}. As a result, many studies rely on human annotators to assess responses along dimensions such as relevance, informativeness, and coherence~\cite{deriu2021survey}. More recent surveys further emphasize the need for multi-dimensional evaluation frameworks that capture interaction quality, user experience, and conversational effectiveness~\cite{marconi2026assessing, guan2026evaluating}.

However, these evaluation approaches are largely developed for open-domain dialogue or customer-facing applications, and do not adequately capture the requirements of expert decision-making contexts. In high-stakes domains such as sustainability, evaluation must go beyond surface-level response quality to consider whether outputs are verifiable, unbiased, contextually grounded, and traceable to primary evidence sources. Studies of domain-specific conversational agents (e.g., healthcare and finance) have begun to introduce expert-informed evaluation dimensions such as trustworthiness, context-awareness, and decision support utility~\cite{10.1145/3706599.3719675, abbasian2024foundation}. Within sustainability-focused conversational systems, evaluation remains particularly underdeveloped. Existing work on conversational agents for environmental sustainability has primarily relied on user studies, questionnaires, or behavioral outcome measures, with limited standardization of evaluation criteria~\cite{sanguinetti2024conversational, giudici2025persuasive}. Even when domain-specific metrics are introduced, they tend to focus on response accuracy or behavioral impact rather than on how experts interpret, verify, and reason with AI-generated knowledge. 

Recent advances in retrieval-augmented generation (RAG) systems have improved factual grounding by combining document retrieval with language model synthesis~\cite{zhu-etal-2019-retrieval}. However, evaluation of these systems still largely centers on correctness or retrieval performance, with limited attention to how users assess credibility, trace evidence, or incorporate system outputs into decision-making workflows. Emerging work on conversational explainable AI further suggests that interaction design and explanation interfaces play a critical role in shaping user trust and reliance in decision contexts~\cite{10.1145/3708359.3712133}.

Taken together, this literature reveals a critical gap: while conversational AI systems are increasingly used in complex, knowledge-intensive domains, evaluation frameworks have not kept pace with the needs of expert users who must critically assess, justify, and act upon AI-generated information. 

Our work addresses this gap by developing a provenance-first conversational prototype and an expert-grounded evaluation framework for sustainability knowledge sensemaking. By incorporating domain-specific evaluation dimensions co-defined with experts, we move beyond generic response quality metrics to better capture how trust, interpretability, and usefulness are assessed in real-world sustainability decision contexts.

\begin{figure}[t!]
    \centering
    \includegraphics[width=0.45\textwidth]{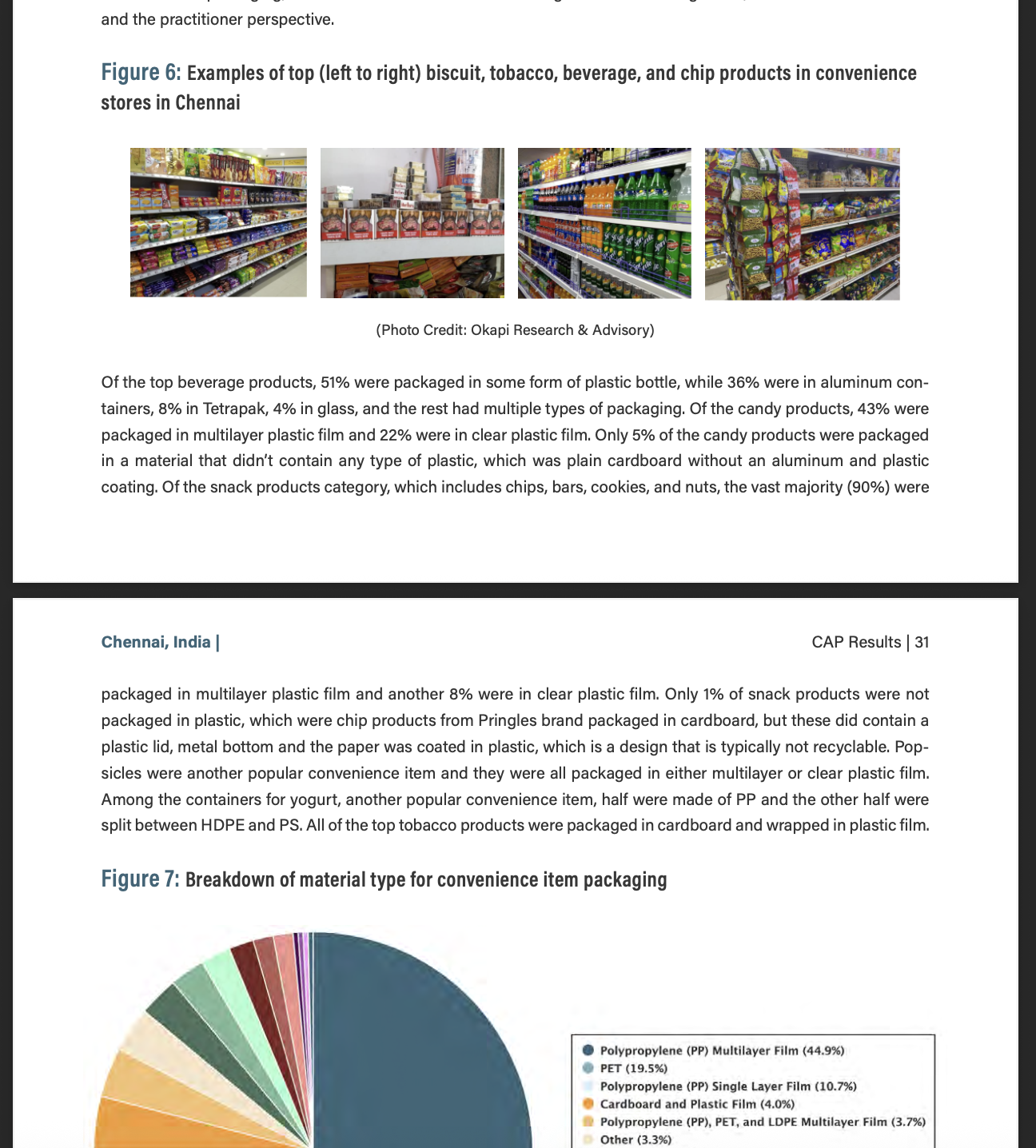}
    \caption{Sample page from a circularity assessment report showing the typical combination of text, image, data visualization, and tables, which makes parsing relevant information from multiple sources a cognitively demanding and time consuming task.}
    \Description{Sample page from a circularity assessment report showing the typical combination of text, image, data visualization, and tables, which makes parsing relevant information from multiple sources a cognitively demanding and time consuming task.}
    \label{fig:sample-page}
\end{figure}

\section{SpheriCity: Design and System Overview}

\textbf{Context.} City governments, environmental organizations, and sustainability researchers increasingly produce circularity assessment reports to evaluate material flows, waste management systems, and policy interventions across urban regions. These reports synthesize information about infrastructure, governance practices, and environmental outcomes. They are widely used to inform sustainability planning and policy development.

While rich in insight, these documents are often long, heterogeneous, and difficult to navigate efficiently. Individual reports frequently exceed 80--100 pages and contain a mixture of technical descriptions, policy analysis, and data tables (a sample page from one such document is presented in Figure~\ref{fig:sample-page}). Experts seeking to compare interventions across cities or identify promising practices must manually search across multiple reports, identify relevant sections, and synthesize evidence from dispersed passages. This process can be time-consuming and cognitively demanding, particularly when experts work under real-world time constraints.

Conversational AI systems offer a potential way to support knowledge exploration across such document collections by enabling natural language interaction with unstructured information sources. However, in sustainability contexts, the ability to verify and interpret AI-generated responses is a critical design criterion. Experts must be able to trace claims back to original sources, assess the scope and limitations of synthesized information, and understand how responses relate to underlying evidence.

\begin{figure}[t!]
    \centering
    \includegraphics[width=0.88\columnwidth]{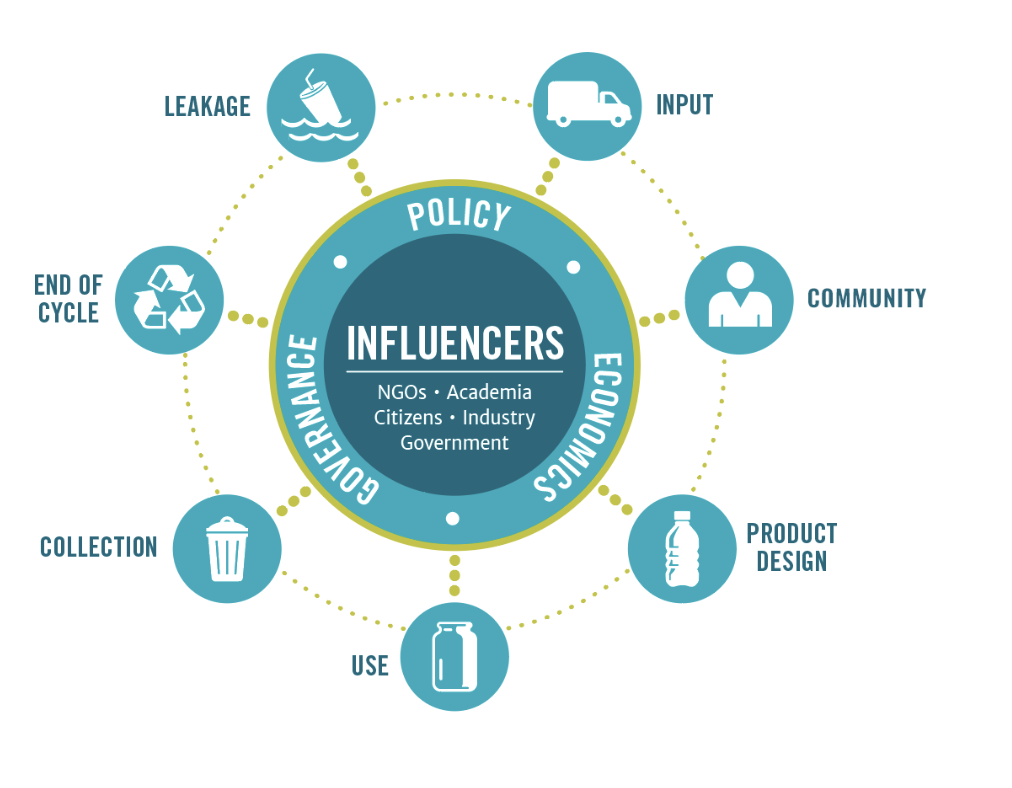}
    \caption{Each Circularity Assessment Report follows a hub-and-spoke model which organizes information around seven key stages of materials flow in an urban environment.}
    \Description{Each Circularity Assessment Report follows a hub-and-spoke model which organizes information around seven key stages of materials flow in an urban environment.}
    \label{fig:hub-model}
\end{figure}

\subsection{Sustainability Report Corpus}

SpheriCity operates over a corpus of city-level circularity assessment reports documenting material flows, waste management practices, policy interventions, and stakeholder perspectives across cities worldwide. Our corpus includes 36 circularity reports, totaling approximately 2,800 pages, with an average report length of about 65 pages. Figure~\ref{fig:hub-model} graphically represents the information flow and organization that takes place at seven stages to generate this report, following the workflow described in ~\cite{undp2023_circular_economy_explainer, maddalene2023circularity, jambeck2024circularity}.

These reports provide valuable domain knowledge but present several interaction challenges. The documents follow semi-structured formats, use inconsistent terminology across cities, and contain information distributed across multiple sections. As a result, answering comparative questions or synthesizing cross-city insights often requires extensive manual reading and cross-referencing. The initial design goal is: rather than requiring users to browse individual PDF documents or rely on keyword searches, SpheriCity enables conversational queries that draw from the entire report corpus while preserving traceability to the original report excerpts that the chatbot used to support generated responses.

\subsection{System Architecture}

To support conversational interaction with the report corpus, we developed \textbf{SpheriCity}, an expert-facing conversational prototype designed to facilitate sustainability knowledge sensemaking. The system uses a retrieval-augmented generation (RAG) architecture~\cite{lewis2020retrieval} that connects natural language queries with relevant passages from the report corpus.

When a user submits a query, the system retrieves relevant report excerpts and supplies them as contextual input to a large language model that synthesizes responses grounded in the retrieved evidence. SpheriCity explores two complementary retrieval strategies: 1) vector-based similarity retrieval and 2) graph-based knowledge retrieval. Vector retrieval identifies semantically similar passages using embedding-based similarity search, while graph retrieval leverages entity relationships extracted from reports to identify connected knowledge paths. 

\begin{figure}[t!]
    \centering
    \includegraphics[width=0.99\columnwidth]{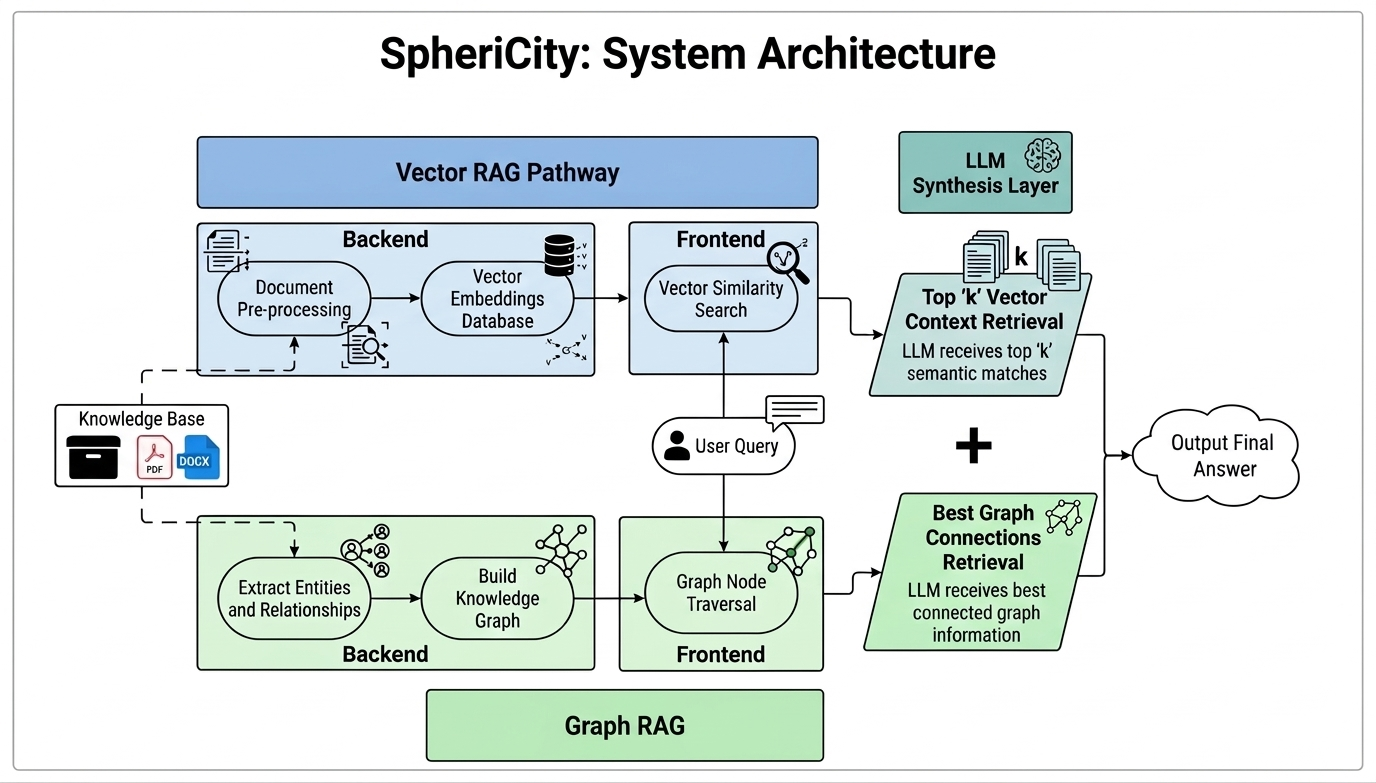}
    \caption{System architecture of SpheriCity.}
    \Description{System architecture of SpheriCity.}
    \label{fig:architecture}

\end{figure}

Retrieved passages are then combined into a context window and passed to the language model for response generation. The generated responses are presented through a provenance-aware interface that includes structured summaries and explicit source citations linking claims to page-level references within the original reports. Figure~\ref{fig:architecture} illustrates the overall system architecture of the SpheriCity conversational system.

\subsection{Response Generation Prompt}\label{Prompt}
\begin{tcolorbox}[colback=gray!5, colframe=black!40, title={System Prompt}]
You are an expert assistant on waste lifecycles, supporting us in developing 
circularity reports. Your role is to synthesize insights from 
multiple retrieval approaches and provide users with a clear, concise, and well-structured 
response to their question.

You will receive two draft answers: one derived from a graph-based retrieval system and one 
derived from a vector-based retrieval system. Your task is to integrate the strongest, most 
well-supported points from both answers into a single reply.

Some rules to follow:

\begin{itemize}
  \item When the two drafts overlap, merge them seamlessly. 
  \item When they diverge, briefly note the discrepancy and prioritize the most reliable evidence.
  \item Cite examples and data from circularity reports with the format \emph{(filename, p.~N)}.
  \item Add qualifiers where uncertainty exists.
\end{itemize}

Your audience is non-technical and so your answer should not make any mention of the underlying architecture that is evidencing your response. 
Ensure headers are separated by line breaks and the final response is accessible, transparent, 
and well-evidenced. Use clear paragraph breaks, bullets, or numbering where helpful.
\end{tcolorbox}

\subsection{Retrieval Configurations}

SpheriCity supports multiple retrieval strategies for grounding response generation. These retrieval architectures determine how relevant evidence is selected from the report corpus before being provided to the language model. We implemented three retrieval configurations that differ in how contextual evidence is constructed:\\

\subsubsection{Vector-based Retrieval.}
\blue{This approach retrieves passages using embedding-based similarity search. We segment each report document into text chunks using a recursive character splitting strategy with a chunk size of 2,500 characters and a 150-character overlap between adjacent chunks. The overlap was implemented specifically to mitigate boundary issues where continuous information might otherwise be split across chunks. We used hierarchical separators (paragraph breaks, sentences, words) during chunking to preserve natural semantic boundaries. Each chunk was encoded using the all-MiniLM-L6-v2 sentence transformer model~\cite{reimers2021allminilml6v2}, which maps text to a 384-dimensional embedding space. We used a FAISS index~\cite{douze2024faiss} with L2 distance for similarity search. Retrieval was implemented as a two-stage process: an initial broad search retrieved k=40 candidate chunks, which were then re-ranked using a cross-encoder model~\cite{crossencoder2021msmarco}. We used cross-encoders because unlike bi-encoders, cross-encoders evaluate query–document pairs jointly, producing more precise relevance scores.} 
\blue{After re-ranking, the top $k=12$ chunks were selected and provided as context to the language model for response generation. The value $k=12$ was determined through pilot testing, which indicated that including more chunks introduced noise without substantial information gain. Retrieved chunks were organized by their source document, with chunk metadata, including page numbers and source document identifiers, preserved throughout to enable accurate source attribution in generated responses.}

\subsubsection{Graph-based Retrieval} This configuration constructs a \blue{knowledge graph from the report corpus by extracting entities and relationships from the report text. Queries trigger traversal of the knowledge graph to identify connected concepts and supporting passages. Retrieved graph-connected evidence is then supplied as context for response generation.}
\blue{We used the LightRAG system~\cite{guo2024lightrag}, which automatically constructs a knowledge graph from document collections by extracting named entities (cities, organizations, materials, policies) and relationships based on co-occurrence, syntactic patterns, and semantic analysis. Entities and their relationships are connected across documents, enabling traversal of multi-hop relationship paths. For each query, graph traversal combined local neighborhood exploration — starting from entities mentioned in the query — with a global context assessment that surfaces important relevant entities not directly mentioned in the query. Retrieval parameters were set to top $k=60$ for entities and chunk top $k=10$ for supplementary vector chunks. The retrieved graph-connected evidence was then supplied as context to the same language model used for vector-based response generation, with an equivalent prompt structure to ensure comparability.}

\subsubsection{Hybrid Retrieval}
\blue{The hybrid configuration combines the outputs of the vector and graph retrieval pipelines to capture both semantically similar passages and conceptually related information across documents into a single synthesized response. We use a \textit{Response Generation Prompt} for creating the final hybrid response.}
\blue{Rather than merging retrieved passages before generation, both pipelines first generate independent responses from their respective retrieved contexts. These two draft responses are then passed together to the language model with an explicit synthesis prompt (see Section~\ref{Prompt}) that instructs the model to: (1) integrate complementary information from both drafts, (2) note discrepancies where the two drafts diverge and prioritize the more evidence-specific claim, and (3) maintain consistent source attribution throughout. This design was motivated by the hypothesis that vector retrieval excels at semantic passage recall while graph retrieval captures structured entity relationships — and that synthesis at the response level, rather than the retrieval level, would better preserve the distinct strengths of each approach. Response generation across all three configurations used GPT-4o-mini from OpenAI~\cite{openai2024gpt4omini} with temperature=0.3 and top-p=0.9, settings chosen to produce factually grounded outputs while retaining sufficient flexibility for coherent text generation.}
\blue{These retrieval strategies serve two purposes. First, they allow us to examine how different forms of evidence grounding influence expert perceptions of trustworthiness, interpretability, and usefulness when interacting with conversational AI systems.}

\subsection{Design Principles}

Beyond retrieval performance, the design of SpheriCity focuses on supporting expert trust, interpretability, and verification needs. Informal interviews with three sustainability experts informed three interaction design principles that guided the development of the interface.

\textbf{Provenance as a Core Interface Element.}
SpheriCity foregrounds evidence provenance within the primary interaction loop. Sustainability experts rarely accept synthesized claims without verification and often need traceable evidence to justify recommendations to stakeholders. By embedding page-level citations directly into each generated response, the interface enables users to trace statements back to the original report context and quickly verify claims.

\blue{A key design decision supporting provenance integrity was the separation of source attribution from language model response generation. Rather than instructing the model to generate citations, page numbers are derived from document metadata preserved during the document preparation pipeline: text was extracted from PDF and Word files using format-specific libraries, with custom pipelines that preserved page numbers, section hierarchies, and document identifiers throughout the chunking and retrieval process. Retrieved chunks carry this metadata forward so that citations in generated responses reference actual retrieved passages rather than being fabricated by the language model. This architectural choice is intended to substantially reduce — though it does not entirely eliminate — the risk of hallucinated page numbers or fabricated document references. However, we acknowledge that the model may still draw on retrieved passages in ways that misrepresent or overgeneralize their content, a risk that motivates our emphasis on visible citations that allow experts to verify claims against the original source text.}

\textbf{Response Structure over Fluency.}
Experts reported that long narrative responses can obscure inconsistencies and make verification more difficult. SpheriCity therefore prioritizes structured bullet-point synthesis that organizes information into interpretable units such as interventions, outcomes, challenges, and recommendations. This structure supports \textit{cognitive offloading} by enabling experts to quickly scan responses, compare cases across cities, and identify missing, uncertain, or conflicting information.

\textbf{Prompt Templates as Interaction Scaffolds.}
In expert-facing contexts, poorly formulated queries can lead to misleading or incomplete answers. To reduce this risk, the interface includes prompt templates that encode common sustainability analysis tasks such as cross-city comparisons, policy summarization, and recommendation generation. These templates guide expert and first-time users alike toward evidence-based information seeking while reducing the cognitive burden of query formulation.

\subsection{User Interface}

The SpheriCity interface (Figure~\ref{fig:chatbot}) consists of three primary components designed to support expert knowledge workflows.

\textbf{Conversational Query Panel.}
Users interact with the system through a chat-style input interface that accepts natural language questions. Domain-specific prompt templates help users formulate structured queries aligned with typical sustainability analysis tasks.

\textbf{Structured Response Display.}
Responses are formatted as structured bullet points rather than free-form text. This organization enables users to quickly identify key insights and compare information across multiple reports without reading long narrative responses.

\textbf{Evidence Provenance Panel.}
Each response is accompanied by explicit source citations, including report titles and page numbers. Users can expand citations to view the original supporting text, enabling rapid verification and traceability to primary evidence.

\section{Expert-Grounded Evaluation}

We approach evaluation as a participatory design activity to identify what sustainability experts care about when making sense of AI-generated insights. Rather than directly using generic conversational agent evaluation metrics and optimizing for task-specific performance, our goal was to co-define domain-specific quality criteria that reflect the real needs of sustainability decision-making. \blue{The experts shaped the evaluation framework and design principles of the conversational agent based on the feedback they provided on our early prototype design.}

\subsection{\blue{Design Workshop} with Sustainability Experts}
We conducted a structured \blue{participatory design workshop} with three sustainability and circularity experts who regularly work with city-level assessment reports for policy analysis, program design, and stakeholder communication. Participants were selected based on their active participation and experience in this domain. This session lasted for about an hour and was conducted over a video-conferencing call, where two of the lead authors were present to facilitate the design conversation.

\begin{table*}[t!] 
\centering 
\caption{Assessment framework with five evaluation dimensions, each measured through three statements.} 
\label{tab:dimensions} 
\begin{tabular}{lp{12cm}} 
\toprule 
\textbf{Dimension} & \textbf{Statements} \\ 
\midrule Topical Relevance & The chatbot used information that was directly relevant to my question. 
\newline The chatbot response stayed on topic and avoided unrelated details. \newline The chatbot addressed all key aspects of my question’s topic. \\ 
\midrule Contextual Relevance & The chatbot understood the intent behind my question. 
\newline The chatbot’s response reflected the specific angle I was asking about. \newline The chatbot considered the context of my question when forming the response. \\ 
\midrule Perceived Accuracy & The chatbot’s response was factually accurate to the best of my knowledge. 
\newline The chatbot avoided including information I believe is incorrect. 
\newline I would trust the information in the chatbot’s response without additional verification. \\ 
\midrule Bias / Neutrality & The chatbot’s response maintained a neutral and unbiased tone. 
\newline The chatbot avoided favoring one perspective over another. 
\newline The response did not promote a specific viewpoint without evidence. \\ 
\midrule Depth of Information & The chatbot answer contained an appropriate level of details for my needs. 
\newline The chatbot avoided giving too much unnecessary detail. \newline The chatbot covered enough information for me to understand the topic fully. \\ 
\bottomrule \end{tabular} 
\end{table*}

During the session, experts were asked to review example chatbot responses generated from an early prototype and to reflect on what made a response useful, trustworthy, or in their professional workflows. Rather than prompting them with predefined metrics, we encouraged open discussion around questions such as: \textit{i) What would make you trust or distrust this answer? ii) What information would you need to verify this claim?
iii) What kinds of errors would be unacceptable in practice?
iv) What level of detail is helpful versus overwhelming?}

These discussions revealed that experts’ judgments were not driven primarily by fluency or completeness, but by deeper concerns about evidence traceability, contextual appropriateness, neutrality, and the risk of misleading generalizations. Several experts explicitly rejected the idea of a single “best” answer in this process, contrary to approaches generally used in evaluating conversational agents' task performance, and rather emphasized that sustainability decisions often involve tradeoffs, contested values, and incomplete data, which informed the design of our current prototype.

\subsection{Can Generic Chatbot Evaluation Metrics Work in Sustainability Contexts?}

Generic chatbot evaluation frameworks typically emphasize dimensions such as fluency, grammatical correctness, relevance, and informativeness. While these criteria are useful for everyday conversational systems, experts argued that they are insufficient and sometimes misleading when applied to sustainability contexts. In particular, experts highlighted several value tensions that are rarely captured by generic metrics:

\textbf{Completeness vs. Hallucination.} More comprehensive answers were often perceived as less trustworthy if they introduced unsupported details or speculative generalizations. Experts preferred partial but verifiable answers over fluent, overly complete ones.

\textbf{Depth vs. Speed.}
While rapid synthesis and summarization are valuable under time pressure, overly summarized responses risk flattening important nuances, caveats, and contextual details. Experts preferred the ability to trade speed for depth depending on the task difficulty.

\textbf{Generalization vs. Contextualization}
Responses that are too generalized across cities or policy contexts were frequently judged as misleading and unhelpful, even if factually accurate in isolation. Experts emphasized that sustainability interventions are deeply context-dependent and must be framed accordingly.

\textbf{Neutrality vs. Persuasiveness.} Experts were wary of answers that appeared to advocate for particular policies repeatedly without acknowledging tradeoffs, uncertainties, or alternative perspectives.
These tensions underscore why general-purpose chatbot evaluation criteria are insufficient in sustainability contexts, which is why we designed evaluation dimensions in collaboration with the sustainability experts that align with their needs and workflow.

\subsection{Expert Co-Defined Evaluation Dimensions}\label{eval}
Through iterative discussion and refinement, we synthesized five domain-specific quality dimensions that reflect how experts actually judge AI-generated sustainability knowledge:
\begin{itemize}
\item \textbf{Topical Relevance.}
Does the response draw information from the appropriate policy or sustainability scope and directly address the user’s question?
\item \textbf{Contextual Fit.}
Is the response appropriately scoped to the geographic, institutional, and policy context of the question?
\item \textbf{Accuracy.}
Are factual claims consistent with the reported evidence, and are key details correctly represented?
\item \textbf{Neutrality.}
Does the response avoid unwarranted skewed perspective, bias, or selective framing of evidence?
\item \textbf{Depth of Information.}
Does the response provide sufficient explanatory detail without overwhelming the user or introducing speculative content?
\end{itemize}

The purpose of these dimensions is to make expert values ingrained in the design of conversational systems, rather than focusing on benchmarking on a single gold-standard answer. For each dimension, the experts formulated three statements for which responses will be evaluated. This resulted in a 15-item evaluation instrument using 5-point Likert scales (1: Strongly Disagree; 5: Strongly Agree). The full evaluation instrument is presented in Table~\ref{tab:dimensions}.



\begin{table*}[t]
\centering
\caption{Representative evaluation queries, grouped by category.}
\label{tab:queries}
\small 
\begin{tabularx}{\textwidth}{lX} 
\toprule
\textbf{Category} & \textbf{Example Query} \\
\midrule
\multirow{5}{*}{\textbf{City-specific}} 
 & $\bullet$ What are the challenges that Ann Arbor faces with compostable plastics and fiber containers? \\
 & $\bullet$ What are some of the substitutes to single-use plastics to optimize circularity that are found in Minneapolis? \\
 & $\bullet$ How might Hawaii's HI-5 beverage container redemption program have influenced the litter found in the environment in Hilo? \\
 & $\bullet$ How does informal recycling impact the litter surveyed in India? \\
\midrule
\multirow{4}{*}{\textbf{Comparison}} 
 & $\bullet$ What are the common barriers to circularity across cities in Southeast Asia? \\
 & $\bullet$ What are some common opportunities among cities with tourism economies? \\
 & $\bullet$ How does access to industrial composting impact a city's overall circularity? \\
 & $\bullet$ How does recycling infrastructure vary in urban versus rural areas? \\
\midrule
\multirow{6}{*}{\textbf{Recommendations}} 
 & $\bullet$ How do the opportunities given by the CIL compare to the opportunities described by interviewees in the Urban Ocean cohort? \\
 & $\bullet$ What projects did cities in the Urban Ocean cohort want to complete as next steps based on opportunities identified during the CAPs? \\
 & $\bullet$ What are some common recommendations given by stakeholders in Manila, Philippines? \\
 & $\bullet$ If EPR for single-use plastics were to be instituted in the cities in India, what could some of the potential outcomes be? \\
 & $\bullet$ What are some opportunities to increase recycling in a small rural city? \\
\bottomrule
\end{tabularx}
\end{table*}

\section{Formative Study}

We conducted a formative expert review to examine how different retrieval strategies influence expert perceptions of conversational AI responses in sustainability knowledge workflows. The goal of the study was not to benchmark system performance, but to understand how different evidence grounding mechanisms affect perceived trustworthiness, interpretability, and usefulness when experts interact with AI-generated knowledge.

\textbf{Participants.} Six sustainability and circularity experts \blue{participated in the study. Participants were recruited through our professional research network based on active domain expertise rather than familiarity with AI systems, reflecting our goal of grounding evaluation in expert values and real-world sustainability workflows. All participants provided informed consent and chose to waive compensation for their time.} 

\textbf{Query Set.}
We curated 13 representative sustainability queries designed to reflect realistic expert information needs when working with circularity assessment reports. The queries spanned several analytical tasks commonly encountered in sustainability decision-making, including cross-city comparison, policy summarization, evidence-based recommendation, and information inquiry. The full list of queries is provided in Table~\ref{tab:queries}.

Queries were selected to highlight known challenges in sustainability knowledge sensemaking, including fragmented evidence across documents, strong contextual dependencies between cities and policy environments, and the risk of misleading generalizations when synthesizing information from multiple reports. We believe the full query set and evaluation instrument used in this study will be useful artifacts to support future research on conversational sustainability systems.

\textbf{Evaluation Design Rationale.} 
We conducted a static expert review rather than a fully interactive multi-turn evaluation. In a static protocol, experts rate pre-generated system outputs rather than interacting with the system directly. This design was chosen for two reasons. First, it substantially reduces evaluation burden: interactive evaluation requires participants to play both user and evaluator simultaneously, which is time-consuming and cognitively demanding for busy domain practitioners~\cite{NEURIPS2019_fc981212}. Second, generating responses in advance allowed us to present all three retrieval configurations for the same query in randomized order within a single session, enabling within-rater, within-item comparisons that would be difficult to achieve through interactive use. We acknowledge that static evaluation trades ecological validity for experimental control — it does not capture the natural conversational dynamics of real information-seeking — and we discuss this limitation in Section~\ref{limitation}.

\textbf{Procedure.} To explore how evidence grounding shapes perceived response quality, we experimented with the three retrieval configurations implemented within SpheriCity:

\begin{itemize}
\item \textbf{Vector Retrieval}: semantic similarity retrieval based on embedding search
\item \textbf{Graph Retrieval}: relational retrieval based on knowledge graph structure
\item \textbf{Hybrid Retrieval}: a combination of semantic and relational retrieval approaches
\end{itemize}

For each query, the system generated three responses corresponding to the three retrieval configurations. Responses were anonymized and presented to participants in randomized order to minimize ordering bias.

Participants evaluated each response using the five expert-defined quality dimensions described in the previous section: topical relevance, contextual fit, perceived accuracy, neutrality, and depth of information. Each dimension was measured using the 15-item Likert-scale instrument shown in Table~\ref{tab:dimensions}.

\begin{figure*}[t!]
    \centering
    \includegraphics[width=0.9\textwidth]{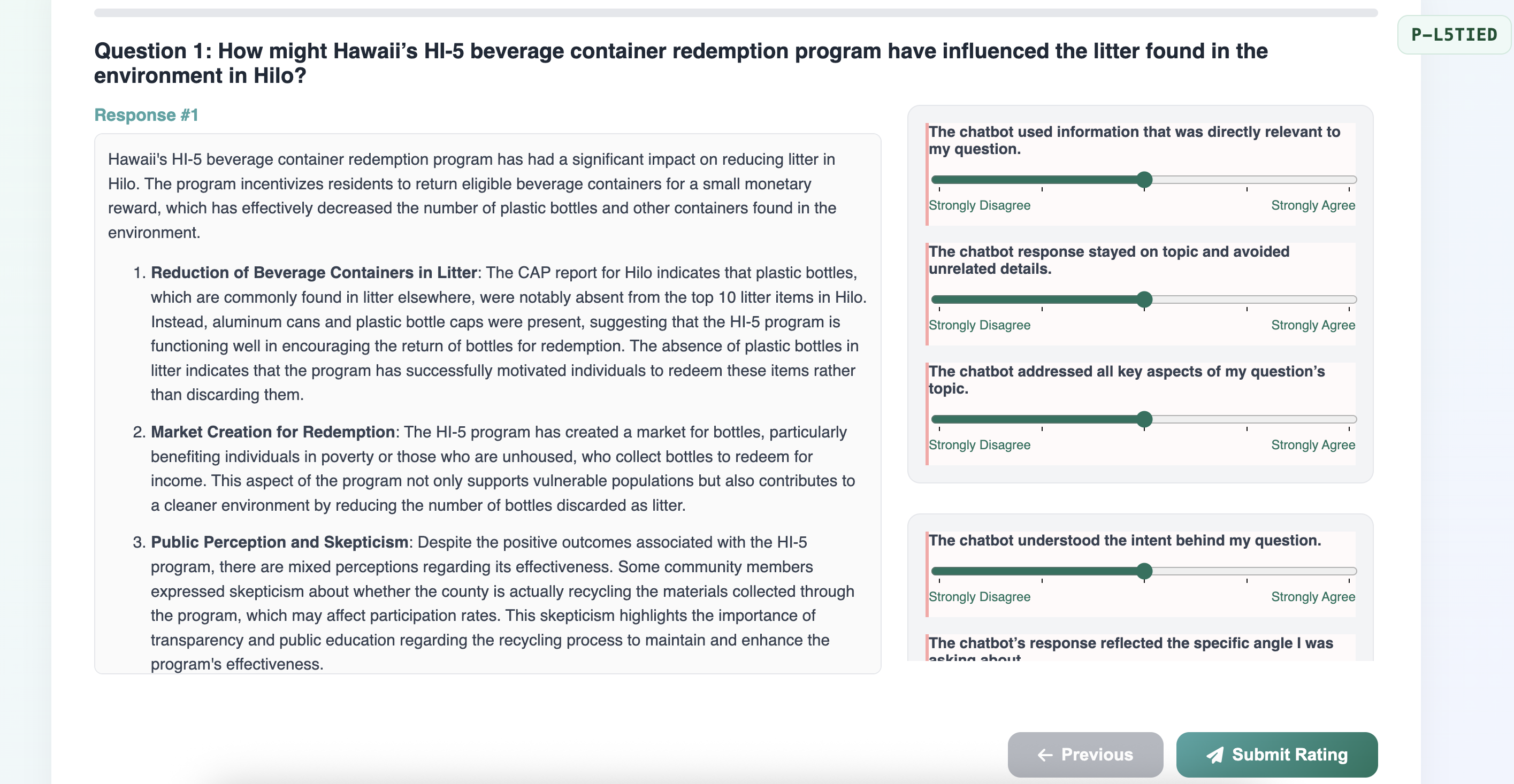}
    \caption{User study interface that displayed responses (left) and the 15 items to be rated on a 5-point Likert scale (right).}
     \Description{User study interface that displayed responses (left) and the 15 items to be rated on a 5-point Likert scale (right).}
    \label{fig:userstudy}
\end{figure*}

Responses were presented through the SpheriCity interface, which supported side-by-side comparison of response variants and provided page-level source citations for evidence verification (see Figure~\ref{fig:userstudy} for an example). Participants were encouraged to reflect on several aspects during evaluation, including:

\begin{itemize}
\item What made a response trustworthy or untrustworthy,
\item Whether supporting evidence appeared sufficient or misleading,
\item Whether the response would be usable in real sustainability decision contexts,
\item How the interface could better support expert workflows.
\end{itemize}

In addition to Likert-scale ratings, participants were invited to provide written comments describing their reasoning and observations. These qualitative reflections were used to supplement the quantitative evaluation results. We acknowledge the small number of experts in our study as a limitation; however, our goal here was not statistical generalization but to surface expert reasoning patterns and interaction concerns that can inform the design of future conversational sustainability systems. 

Each evaluation session was conducted individually and lasted approximately 2 to 2.5 hours. The study protocol was reviewed and approved by the Institutional Review Board (IRB) at the authors' institution under expedited review.

\section{Findings}

We analyze the formative study using a mixed-method approach combining qualitative feedback from expert participants with quantitative ratings collected through the expert-grounded evaluation instrument described in the previous section. Together, these findings reveal how sustainability experts interpret conversational AI responses when performing knowledge sensemaking tasks.

\subsection{Qualitative Findings}\label{qualitative}

We analyzed open-ended expert feedback using inductive thematic analysis. The two lead authors independently reviewed participant comments and iteratively grouped observations into themes reflecting how experts judge conversational AI responses in sustainability knowledge work. Through discussion and consolidation, four recurring themes emerged that capture how experts interpret trustworthiness, interpretability, and usefulness in chatbot-assisted knowledge sensemaking.

Table~\ref{tab:themes} summarizes these themes and their implications for conversational system design.\\

\textbf{Theme 1: Trust Requires Provenance}

Across participants, visible source attribution emerged as a prerequisite for trust. Responses that referenced specific reports, cities, and page numbers were consistently perceived as more credible and actionable. Conversely, responses lacking traceable evidence were treated with skepticism, even when they appeared fluent or coherent.

One expert noted that they appreciated answers that included ``specific examples from CAP reports and citations to particular reports to back up claims" (P1). Another emphasized that responses were most trustworthy when they ``directly referred to the report and page number" (P2).

Importantly, provenance influenced not only perceived accuracy but also willingness to use the system in practice. Participants emphasized that synthesized content was only useful when they could verify the underlying evidence quickly. As one expert explained, the system should provide ``the sources that contain the relevant information so users can dive deeper if needed" (P1).

\textbf{Design implication.} In sustainability decision contexts, trust is not inferred from response fluency alone. Instead, it is calibrated through visible evidence grounding that enables experts to trace synthesized claims back to primary sources. \blue{This finding aligns with a broader pattern in expert knowledge work: sustainability practitioners are trained to trace claims to primary evidence as part of their professional practice~\cite{muller2019datascience}. An automated system that omits this affordance does not simply fail a usability criterion, it violates a professional norm, which explains why the absence of citations was treated as a credibility disqualifier rather than a minor inconvenience.}\\

\begin{table*}[t!]
\centering
\caption{Summary of qualitative themes identified from expert feedback.}
\label{tab:themes}
\begin{tabular}{p{5.7cm} | p{8cm}}
\toprule
\textbf{Theme} & \textbf{Key Insight} \\
\midrule
{Trust Requires Provenance} & {Experts rely on visible citations and evidence traceability to evaluate credibility.} \\
Sensemaking Over Answer Accuracy & Experts value structured synthesis that supports reasoning rather than single definitive answers. \\
Interpretability Breakdowns Trigger Distrust & Contextual or geographic errors quickly undermine confidence in the system. \\
Expert–LLM Workflow Misalignment & Chatbot outputs serve as starting points but require additional tools to support real decision workflows. \\
\bottomrule
\end{tabular}
\end{table*}

\textbf{Theme 2: Experts Value Sensemaking Support Over Definitive Answers}

Experts did not judge response quality solely in terms of factual correctness. Instead, they valued responses that supported knowledge sensemaking through structured synthesis, contextual framing, and examples that enable deeper reasoning.

One participant described the responses as ``useful and accurate… for a summary or introduction to an issue," but emphasized that their usefulness depended heavily on how the information was structured and contextualized. Structured bullet-point responses were consistently preferred over long narrative paragraphs because they were perceived as easier to scan, compare, and verify.

Participants also emphasized that conversational systems should support exploration rather than presenting single definitive answers. One expert noted that ``it’s hard to capture everything in one response," suggesting that recommended reading features would better align with expert workflows.

\textbf{Design implication.} Conversational sustainability tools should prioritize synthesis and exploration rather than presenting responses as complete answers. This preference reflects a fundamental characteristic of sustainability decision-making: decision making answers are rarely well-defined and singular, and expert value lies in navigating trade-offs rather than receiving exact recommendations. Designing for sensemaking rather than question answering aligns with prior work on human–AI collaboration that emphasizes supporting iterative reasoning over automating conclusions~\cite{amershi2019guidelines, lim2009intelligibility}.\\

\textbf{Theme 3: Interpretability Breakdowns Trigger Distrust}

Participants identified several cases where geographic or institutional references were mis-scoped or incorrectly attributed. These errors immediately undermined confidence in the system’s outputs.

One expert described multiple contextual inconsistencies:

``Some responses referred to Florida as a city even though Florida is a state… there were also mentions of the Athens report for Urban Ocean questions even though Athens is not part of that cohort… Panama City, Panama was referenced in the Southeast Asia question even though it is in Central America" (P5).

Another participant noted that the chatbot occasionally ``discussed the circularity reports for Florida as if it were a city" (P6). These mistakes were interpreted not as minor errors but as signals that the system might be overgeneralizing or misrepresenting report content.

\textbf{Design implication.} Even small contextual errors can disproportionately erode trust in high-stakes domains. Conversational systems should therefore surface uncertainty and clearly communicate contextual limitations when evidence is incomplete. This disproportionate impact of small contextual errors on trust is consistent with findings from high-stakes human–AI interaction research, where calibrated trust depends on the system demonstrating accurate self-knowledge of its limitations~\cite{amershi2019guidelines}. In sustainability contexts, where experts are responsible for the downstream consequences of decisions, a single factual error — such as a misidentified geographic scope — is sufficient to disqualify an entire response, regardless of the quality of surrounding content of that response.\\

\textbf{Theme 4: Expert–LLM Workflow Misalignment}

Although experts were generally impressed with the system’s capabilities, they also identified misalignments between chatbot outputs and real decision-making workflows. Responses were perceived as useful starting points for exploration but not yet ready for direct integration into policy reports or recommendations.

One expert described the responses as valuable for ``general knowledge of an issue," but noted that the appropriate level of detail depended heavily on the intended use case, such as educational materials versus policy recommendations.

Participants also expressed interest in features that would support reflective reasoning during evaluation. For example, one expert suggested that the interface should allow users to ``add notes about why I answered the way I did or what issues I saw in the responses" (P3).

\textbf{Design implication.} Conversational systems supporting expert decision-making should incorporate tools for annotation, evidence tracking, and justification rather than treating responses as final outputs. This misalignment is consistent with prior work framing AI systems as `generative artificial experts' whose value lies in augmenting rather than replacing human judgment in knowledge work~\cite{sowa2025expert}. The experts in our study intuitively positioned the SpheriCity chatbot in this role, but the current interface did not provide the annotation, note-taking, or evidence-tracking features that would support that collaborative mode.

\subsection{Quantitative Findings}

In addition to qualitative feedback, experts provided numerical ratings along the five expert-defined evaluation dimensions: topical relevance, contextual fit, perceived accuracy, neutrality, and depth of information. The evaluation generated a rich dataset comprising 3,510 individual ratings (6 specialists × 13 queries × 3 retrieval
methods × 15 assessment questions). 

With 6 experts rating 3 system responses for
each of 13 questions (within-rater, within-item design), we perform linear mixed-effects analysis to leverage 78 paired
observations and compare each two-system pair. Retrieval configuration was treated as a fixed effect, while participant and query were modeled as random effects to account for repeated measurements within experts and tasks. We calculated the score for each dimension by
averaging the scores for the three individual questions within each dimension. By averaging
items within each dimension, we mitigate possible low-power issues due to our small sample size. All reported differences are on the 1–5 scale. Holm–Bonferroni corrections were applied within each contrast set to adjust $p$-values and reduce the false discovery rate.

The purpose of comparing retrieval configurations was not to benchmark retrieval performance but to probe how different evidence-grounding strategies influence expert perceptions of trust, interpretability, and response usefulness.  \blue{The goal of the quantitative analysis below is to characterize the magnitude and consistency of differences in expert ratings, providing quantitative support to the qualitative themes identified in Section~\ref{qualitative}. Note that given the limited sample size, we interpret these results as directional patterns providing descriptive support for the qualitative analysis rather than definitive statistical comparisons leading to causal conclusions.}

\begin{figure}[t!]
\centering
\includegraphics[width=0.45\textwidth]{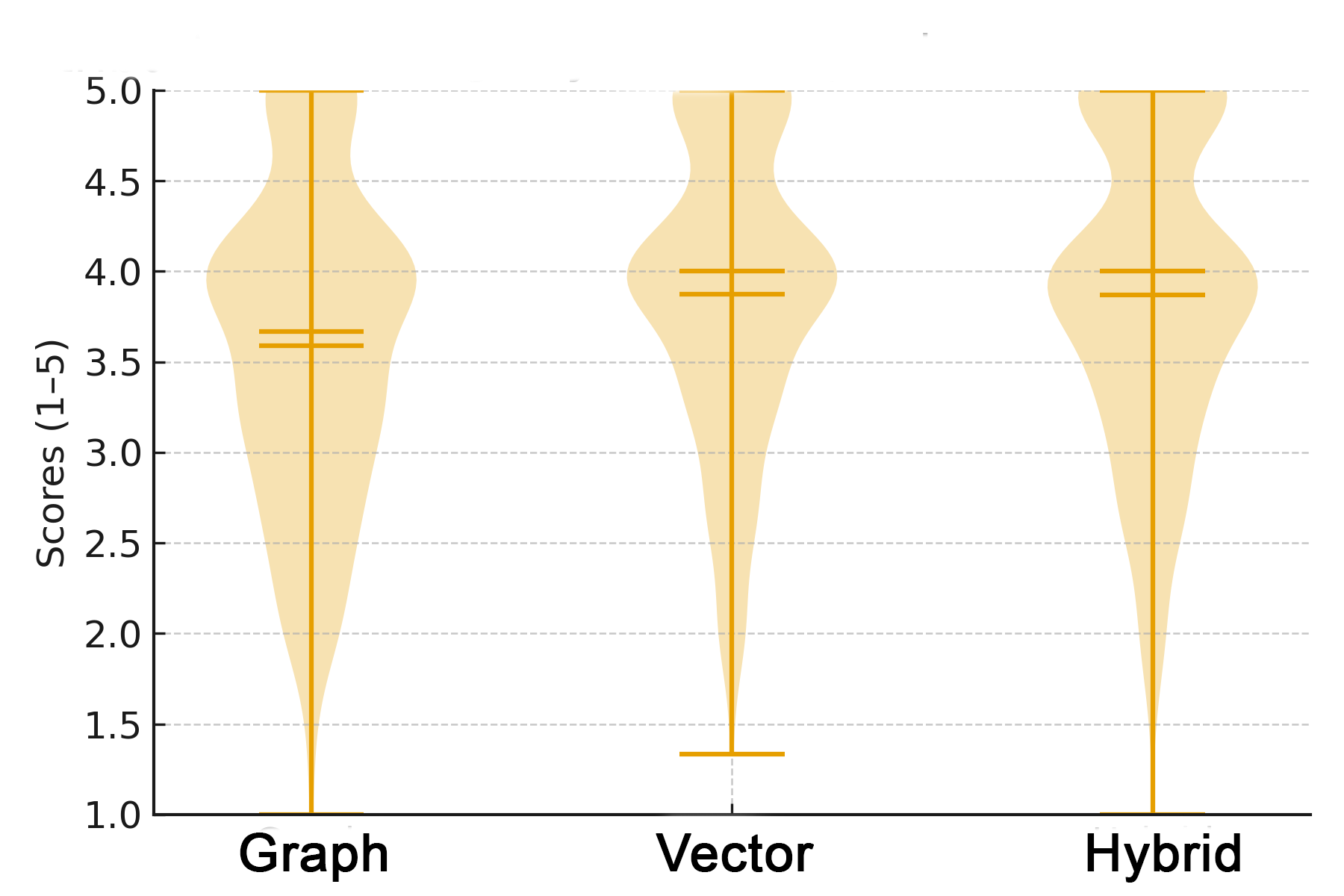}
\caption{Distribution of pooled ratings by system. The violin plot shows that ratings for Graph tend to be lower, while Vector ratings shift upward.}
\Description{Distribution of pooled ratings by system. The violin plot shows that ratings for Graph tend to be lower, while Vector ratings shift upward.}
\label{fig:rq1-violin}
\end{figure}

Figure~\ref{fig:rq1-violin} illustrates the overall distribution of ratings across the three retrieval configurations. Responses generated using vector-based and hybrid retrieval were generally rated higher than those generated using graph-based retrieval. Ratings for graph retrieval were concentrated at lower values and exhibited greater variance, while vector and hybrid responses clustered more strongly toward the upper end of the scale.

\textbf{Overall patterns.}
Across queries, vector-based and hybrid retrieval approaches tended to produce responses that experts perceived as deeper and more accurate than those generated using graph-based retrieval. However, qualitative feedback revealed that these same responses sometimes introduced subtle hallucinations or contextual overgeneralizations, reinforcing the qualitative tension between completeness and trustworthiness observed in Theme~3.

\textbf{Performance across task types.}
We also examined whether system performance varied across different query types, including city-specific questions, cross-city comparisons, and recommendation-oriented tasks. As shown in Table~\ref{tab:rq3}, vector-based and hybrid retrieval consistently outperformed graph retrieval across all task categories. The largest difference occurred for city-specific queries, suggesting that vector-based retrieval may better capture localized contextual information when synthesizing responses.

\begin{table*}[t]
\centering
\caption{System contrasts by task type (adjusted mean differences on a 1--5 scale).}
\label{tab:rq3}
\begin{tabular}{lccc}
\toprule
Task type & Vector--Graph & Hybrid--Graph & Hybrid--Vector \\
\midrule
City-specific   & +0.31, $p<.01$ & +0.29, $p<.01$ & n.s. \\
Comparison      & +0.29, $p<.01$ & +0.28, $p<.01$ & n.s. \\
Recommendation  & +0.27, $p<.05$ & +0.30, $p<.05$ & n.s. \\
\bottomrule
\end{tabular}
\end{table*}

\textbf{Dimension-specific effects.}
Table~\ref{tab:rq4} presents system differences across the five expert-defined evaluation dimensions. The largest improvements for vector and hybrid retrieval appear in perceived depth of information and perceived accuracy. These results suggest that embedding-based retrieval helps the system synthesize more comprehensive responses from the sustainability reports.

However, improvements in completeness did not always translate into higher trust. As reflected in the qualitative findings, experts were particularly sensitive to contextual errors or unsupported generalizations. This highlights the importance of balancing response depth with interpretability and evidence traceability.

\begin{table*}[t]
\centering
\caption{System differences by evaluation dimension (adjusted mean differences on 1--5 scale).}
\label{tab:rq4}
\begin{tabular}{lccc}
\toprule
Dimension & Vector--Graph & Hybrid--Graph & Hybrid--Vector \\
\midrule
Topical Relevance    & +0.23, $p = .036$ & +0.23, $p = .036$ & n.s. \\
Contextual Relevance & +0.18, n.s.       & +0.21, n.s.       & n.s. \\
Perceived Accuracy   & +0.33, $p = .004$ & +0.34, $p = .004$ & n.s. \\
Bias / Neutrality    & +0.28, $p = .009$ & +0.28, $p = .009$ & n.s. \\
Depth of Information & +0.42, $p < .001$ & +0.42, $p < .001$ & n.s. \\
\bottomrule
\end{tabular}
\end{table*}

Taken together, these results suggest that retrieval architecture influences how experts evaluate conversational responses in sustainability contexts. While vector-based retrieval improves response depth and perceived accuracy, the qualitative findings highlight that trust ultimately depends on interpretability and provenance rather than response completeness alone. These findings suggest that different retrieval strategies may support different types of sustainability knowledge tasks rather than indicating a universally superior architecture.

\section{Discussion}

Our findings highlight important challenges and opportunities for conversational AI systems intended to support sustainability decision-making. While recent advances in large language models enable rapid synthesis of information across large document collections, our study shows that the usefulness of such systems in sustainability contexts depends less on response fluency and more on how well they support expert knowledge sensemaking. These observations resonate with prior work in sustainable HCI, which emphasizes the importance of designing technologies that support accountability, long-term reasoning, and value-sensitive decision-making in complex socio-environmental systems.

\subsection{Conversational AI for Sustainability Knowledge Work}

Sustainability experts frequently work with fragmented knowledge sources, including policy reports, environmental assessments, and stakeholder documentation. These documents often contain rich insights but require significant effort to navigate and synthesize. Our findings suggest that conversational AI systems can support this work by lowering the interaction cost of exploring large document collections, consistent with prior calls for human-centric AI approaches to sustainable development~\cite{how2020humancentric}.

However, experts did not treat chatbot responses as definitive answers. Instead, they used the system as an exploratory tool for identifying relevant evidence, understanding cross-city patterns, and locating supporting examples within reports. \blue{This mode of use aligns with prior work on expert sensemaking in data science and knowledge work which emphasizes iterative exploration rather than single-answer retrieval~\cite{muller2019datascience}. Our findings extend the characterization to sustainability knowledge work specifically, where the added dimension of policy accountability — that experts must justify recommendations to stakeholders — makes the traceability of evidence a first-order concern rather than a secondary feature.}

\blue{These findings suggest that conversational sustainability tools should be designed primarily as \textit{sensemaking interfaces} rather than automated decision-support systems. Systems that emphasize exploration, contextualization, and evidence verification may better align with how sustainability professionals interpret complex environmental knowledge. Prior work in sustainable HCI has emphasized that computational tools in this domain must support accountability and long-term reasoning~\cite{knowles2013sustainability, disalvo2010sustainable, 10.1145/2493431} — our findings add that these values manifest concretely in the design features experts rely on: visible citations, structured synthesis, and contextual scoping of the query.}

\subsection{Designing Trustworthy AI for High-Stakes Domains}

Our study also reveals that trust in conversational systems is shaped by factors that differ substantially from typical chatbot evaluation criteria. While many conversational AI benchmarks emphasize fluency, coherence, and response completeness~\cite{deriu2021survey, liu2016evaluate}, experts in our study focused on evidence traceability, contextual accuracy, and interpretability, a gap that mirrors concerns raised in domain-specific evaluation literature for healthcare and other high-stakes contexts~\cite{abbasian2024foundation, 10.1145/3706599.3719675}.

Visible provenance emerged as a key design requirement. Experts consistently reported that responses accompanied by explicit source citations and page-level references were significantly more trustworthy than responses without clear evidence grounding. This finding supports prior work in human–AI interaction emphasizing the importance of intelligible and transparent system outputs~\cite{lim2009intelligibility}. Additionally, this provenance element influenced not only perceived accuracy but also willingness to use the system in practice, suggesting that trust in this context is instrumental rather than affective: experts trusted what they could verify instead of what sounded confident or accurate.

\blue{At the same time, our findings surface a tension that has not been fully articulated in prior work on conversational AI for sustainability: retrieval approaches that produced more detailed, comprehensive answers were sometimes *less* trusted when they introduced subtle contextual errors or overgeneralizations. This inverts the typical assumption that more information is better. In sustainability contexts, where decisions must be justified and contextualized, even small inaccuracies can disqualify an otherwise useful response. This finding aligns with work on trust calibration in human–AI interaction, which emphasizes that systems should surface uncertainty and limitations rather than projecting false completeness~\cite{amershi2019guidelines}.} 

These results suggest that conversational systems designed for sustainability domains should prioritize trust calibration mechanisms such as provenance, uncertainty signaling, and context-sensitive explanations rather than optimizing purely for response completeness.

\subsection{Implications for Human–AI Collaboration in Sustainability}

Experts consistently positioned the chatbot as an assistant rather than an authority. Rather than relying on the system to make recommendations directly, participants used responses as starting points for deeper investigation and evidence gathering.

This observation has implications for how AI systems should be integrated into sustainability decision-making workflows. Instead of replacing expert reasoning, conversational systems should support collaborative sensemaking processes that combine AI-driven synthesis with human judgment and contextual expertise.

\blue{Design features that support human–AI collaboration — traceability, annotation support, suggested reading, multi-step exploration — appear more valuable to experts than features that optimize response completeness. Prior work on conversational decision-support systems supports this direction, showing that interactive evidence presentation improves users' ability to interpret information and make informed decisions~\cite{10.1145/3485447.3512248}. In our context, experts specifically requested the ability to annotate responses, add notes about their reasoning, and flag issues — features that would position the system as a collaborative workspace rather than a query-response interface. Supporting this form of human–AI collaboration may be particularly important in sustainability contexts where decisions involve trade-offs between environmental, economic, and social considerations that no current AI system can resolve unilaterally.}

\subsection{Design Implications for Trustworthy Conversational Systems}

Drawing on the integration of our qualitative and quantitative findings with prior work, we identify four design implications for conversational systems intended for expert-facing sustainability contexts.

\textbf{Design for Active Trust Calibration, Not Fluency.} Trust in our study was not inferred from response fluency. It was earned through visible evidence grounding: specific citations, page references, and explicit scope indicators. This extends Amershi et al.'s guidelines for human-AI interaction in ~\cite{amershi2019guidelines} — specifically the principle that systems should support appropriate reliance — to sustainability knowledge contexts, where the stakes of misplaced trust are consequential for policy and environmental outcomes. Systems should expose the provenance and scope of each claim rather than presenting synthesized content as uniformly authoritative.

\textbf{Support Knowledge Sensemaking Rather Than Answer Generation.} Experts require tools that support synthesis, comparison, and exploration across multiple sources rather than single definitive answers. This aligns with the iterative, verification-oriented workflows described in expert knowledge work research~\cite{muller2019datascience} and suggests that features like suggested reading, follow-up query scaffolds, and side-by-side comparison are higher-value investments than improvements to response fluency.

\textbf{Communicate Uncertainty and Contextual Constraints Explicitly.} Even small contextual errors — misidentified geographies, incorrectly attributed cohorts — disproportionately eroded trust in our study. Systems should therefore surface uncertainty visibly, using explicit qualifiers when synthesizing across heterogeneous sources and clearly marking the geographic or institutional scope of each claim. Conversational explainable AI research further supports this direction, showing that explanation interfaces play a critical role in shaping appropriate reliance in decision contexts~\cite{10.1145/3708359.3712133}.

\textbf{Design for Human–AI Collaboration with Annotation and Verification Support.} Experts consistently positioned chatbot outputs as starting points rather than final products. Future iterations should incorporate tools for response annotation, issue flagging, and justification logging that support expert reasoning and accountability. This positions the interface as a collaborative workspace rather than a query endpoint — consistent with calls for AI systems that support human oversight in high-stakes domains~\cite{amershi2019guidelines, sowa2025expert}. 

\subsection{Limitations and Future Work}\label{limitation}

This work has several limitations that point toward important directions for future research.

First, our expert sample was relatively small and domain-specific. While appropriate for an exploratory design study, larger and more diverse expert populations would allow future work to examine how evaluation criteria vary across sustainability roles and institutional contexts. Second, our study evaluated responses in a static review setting rather than through fully interactive multi-turn conversations. Real-world sustainability analysis often involves iterative questioning and evolving hypotheses. Future work should therefore investigate how trust and interpretability evolve during longer conversational interactions with sustainability knowledge systems. Third, the corpus and queries used in this study were restricted to circularity assessment reports. Sustainability decision-making encompasses many additional domains, including climate adaptation, biodiversity conservation, and environmental justice. Future work should explore whether the design principles identified here generalize across these broader sustainability knowledge contexts. Finally, we did not observe experts using the system during real decision-making episodes. Studying conversational sustainability tools during authentic policy or planning workflows would provide stronger evidence about their practical impact.

\blue{Our study also did not systematically measure hallucination rates, instances where the system generated responses that sounded credible but contained inaccurate claims, including potentially fabricated page numbers or misattributed report content. While our metadata-preserving extraction pipeline structurally reduces the risk of fabricated citations, the language model may still misrepresent or overgeneralize the content of retrieved passages. Expert qualitative feedback identified several such cases (see Theme 3 in Section~\ref{qualitative}). Future work should develop systematic hallucination evaluation protocols appropriate for domain-specific sustainability corpora, including annotation of contextual errors, geographic misattributions, and unsupported generalizations — failure modes that our current evaluation framework captures only through qualitative observation.}

Rather than viewing these limitations solely as constraints, we interpret them as open research questions for the computing and sustainability community:

\begin{itemize}
\item What does trustworthy AI look like when incorrect answers can have civic or environmental consequences?
\item How should uncertainty be represented in sustainability decision-support tools?
\item When should AI systems defer to human judgment rather than produce a response?
\item How can conversational systems better support evidence justification and accountability?
\end{itemize}

Addressing these questions will require continued collaboration between AI researchers, sustainability practitioners, and human–computer interaction scholars.

\section{Conclusion}

Sustainability decision-making increasingly depends on navigating large, fragmented collections of environmental reports, policy documents, and technical assessments. While conversational AI systems promise to accelerate knowledge access and synthesis, their usefulness in high-stakes sustainability contexts depends not only on response quality but also on how well they support expert reasoning, verification, and trust calibration. In this work, we presented \textbf{SpheriCity}, an expert-grounded conversational prototype designed to support knowledge sensemaking across circularity assessment reports. Through a participatory design process and formative expert evaluation, we identified key factors shaping expert trust in AI-generated responses, including evidence provenance, contextual grounding, interpretability, and workflow alignment. Our findings show that sustainability experts treat conversational systems not as authoritative decision-makers but as exploratory tools for locating evidence and supporting reasoning. These insights suggest that the future of AI in sustainability lies not in replacing expert judgment, but in designing systems that support transparent, accountable, and collaborative human–AI knowledge work. By foregrounding provenance, interpretability, and expert workflows, conversational systems can better contribute to responsible and trustworthy sustainability decision-making.



\begin{acks}

This work is supported by NSF Convergence Accelerator Track I: SpheriCity -- Circularity from Molecules to the Built Environment in Communities (Award No: 2345080). We thank the anonymous reviewers for their time and feedback to improve this work.


\end{acks}

\bibliographystyle{ACM-Reference-Format}
\bibliography{references}

\appendix

\begin{table}[t!]
\centering
\caption{Cities with circularity reports, grouped by region.}
\label{tab:cap-cities}
\footnotesize
\begin{tabular}{p{0.30\columnwidth} p{0.65\columnwidth}}
\toprule
\textbf{Region} & \textbf{City} \\
\midrule
\multirow{17}{*}{North America}
  & Ann Arbor (USA) \\
  & Athens, Georgia (USA) \\
  & Atlanta (USA) \\
  & Blytheville (USA) \\
  & Cape Girardeau (USA) \\
  & Tifton (USA) \\
  & Georgetown County (USA) \\
  & Cherokee County (USA) \\
  & Florida Keys (USA) \\
  & Hilo, Hawaii (USA) \\
  & Jekyll Island (USA) \\
  & Miami (USA) \\
  & Minneapolis (USA) \\
  & Pittsburgh (USA) \\
  & San Antonio (USA) \\
  & Santa Fe (USA) \\
  & Vicksburg (USA) \\
\midrule
\multirow{2}{*}{Caribbean}
  & Aruba \\
  & Dominica \\
\midrule
Central America & Panama City (Panama) \\
\midrule
\multirow{2}{*}{South America}
  & Salvador (Brazil) \\
  & Santiago (Chile) \\
\midrule
\multirow{4}{*}{South Asia}
  & Chennai (India) \\
  & Mumbai (India) \\
  & Pune (India) \\
  & Maldives \\
\midrule
\multirow{8}{*}{Southeast Asia} 
  & Bangkok (Thailand) \\
  & Can Tho (Vietnam) \\
  & Hanoi (Vietnam) \\
  & Nam Dinh (Vietnam) \\
  & Melaka (Malaysia) \\
  & Manila (Philippines) \\
  & Semarang (Indonesia) \\
  & Timor Leste \\
\midrule
\multirow{2}{*}{East Africa}
  & Dar es Salaam (Tanzania) \\
  & Seychelles \\
\bottomrule
\end{tabular}
\end{table}
\end{document}